# Morphological evolution in a strained-heteroepitaxial solid droplet on a rigid substrate: Dynamical simulations

Tarik Omer Ogurtani,[1] Aytac Celik[2] and Ersin Emre Oren[3]

*Department of Metallurgical and Materials Engineering, Middle East Technical University, 06531, Ankara, Turkey*

## ABSTRACT

A systematic study based on the self-consistent dynamical simulations is presented for the spontaneous evolution of flat solid droplets (bumps), which are driven by the surface drift diffusion induced by the capillary and mismatch stresses, during the development of the Stranski-Krastanow island morphology on a rigid substrate. The physico-mathematical model, which bases on the irreversible thermodynamics treatment of surfaces and interfaces with singularities (Ogurtani, T.O., J. Chem. Phys. **124**, 144706, 2006) furnishes us to have auto-control on the otherwise free-motion of the triple junction contour line between the substrate and the droplet without presuming any equilibrium dihedral contract (wetting) angles at the edges. During the development of Stranski-Krastanow islands through the mass accumulation at the central region of the droplet via surface drift-diffusion with and/or without growth, the formation of an extremely thin wetting layer is observed. This wetting layer has a thickness of a fraction of a nanometer and covers not only the initial computation domain but also its further extension beyond the original boundaries. Above a certain threshold level of the mismatch strain and/or the size (i.e. volume) of the droplets, which depends on the initial physicochemical data and the aspect ratio (i.e., shape) of the original droplet, the formation of the multiple islands separated by the shallow wetting layers is also observed. By continuously front-tracking the developments in the peak height, in the extension of the wetting layer beyond the domain boundaries, and the change in triple junction contact angle, we observed that these quantities are reaching certain saturation limits or plateaus, when the growth mode is turned-off. Therefore, according to the accepted irreversible thermodynamic terminology as coined by celebrated Prigogine, we state that the Stranski-Krastanow island morphologies are genuine stationary non-equilibrium states. Our theory allows us to observe the dynamical behavior of Stranski-Krastanow island formation without even introducing any external perturbations such as Sine wave undulations or white noise on the original surfaces of the droplets.

---

[1] Corresponding author, Tel.: +90-312-210-2512; fax: +90-312-210-1267; e-mail: ogurtani@metu.edu.tr
Url: http://www.csl.mete.metu.edu.tr

[2] Electronic mail: e104548@metu.edu.tr

[3] Electronic mail: eeoren@uw.edu

# I. INTRODUCTION

The Stranski-Krastanow (SK) morphology, i.e. formation of 'nanoscale islands' or 'quantum dots (QDs)' separated by a thin flat wetting layer, is a general growth characteristics observed in many epitaxially-strained thin solid film systems such as in [In$_x$Ga$_{1-x}$ As/GaAs][1] and [Ge/Si].[2] The formation of quantum dots through the SK growth mode on epitaxially strained thin film surfaces has attracted great attention in the last two decades due to unique electronic and optical properties of QDs.[3,4] The fundamental understanding of the SK growth mode will provide insights necessary to control precise positioning of QDs and may open new avenues in QDs fabrication techniques. It was shown so far that various parameters including surface energy anisotropy, strain level, wetting conditions and growth kinetics, affect how the surface evolution would reach a prescribed stationary state i.e. SK morphology. In general, nonlinear analyses in two dimensional configurations showed that the stress-driven surface instabilities evolve into deep, crack like groove or cusp morphologies.[5,6] However, unlike the semi-infinite homogenous solids, the presence of a substrate affects the instabilities in several ways: First, a stiffer substrate tends to stabilize the film and increases the critical wave length, while the opposite is true for softer substrates. At the limit of a rigid substrate, a critical film thickness exists as shown by Spencer *et al.*,[7,8] below which the film is stable against perturbations of any wave lengths. Moreover, the existence of an interface between the film and the substrate brings more complexity to the problem. At the close proximity of the film surface to substrate, short range wetting interactions dominate and cause an increase in the local surface free energy of the film. This increase hinders the penetration of islands through the boundary layer, and thereby avoids the formation of the Volmer-Weber (VW), i.e. island formation, type of growth mode and promote SK morphology by preventing the surface of the substrate between islands to exposure to the immediate environment.[9,10,11]

The growth or formation of islands in epitaxially thin solid films is a subset of a more general problem namely capillary- and stress-driven shape and microstructural evolution in solids. Asaro and Tiller[12] made the first serious attempt to develop an equilibrium thermostatic model of interfacial morphology evolutions during stress corrosion cracking by adding the elastic strain energy density (ESED) to the so-called chemical potential employed in their paper (see Ref. [12]). The Asaro/Tiller (AT) theory has shown some partial success for the isochoric systems, where the elastic strain energy density enters correctly into the scenario with a positive sign in the Helmholtz free energy density. Grinfeld[13] utilized the Gibbs-Duhem stability theory of thermodynamic equilibrium for the isothermal and isochoric systems, characterized by the second variance in the total Helmholtz free energy denoted as $\delta^2 F > 0$ for the infinitesimal perturbations on the surface morphology associated with the *surface acoustic waves* generated in the nonhydrostatically stressed linear elastic solids in contact with their melts. Freund and Jonsdottir[11] employed the same criterion and developed the instability theory of a biaxially stressed thin film on as substrate due to material diffusion over its free surface by considering only the surface and the elastic strain energies associated with thin film and the substrate, respectively. In all these theories cited there are two premises in common; they all treated the *isochoric* systems implicitly or explicitly, and they all predict that there is a *critical wave length* above which the flat free surface becomes unstable under the sinusoidal perturbations if certain conditions prevail.[14]

Spencer[9] and Tekalign and Spencer[15,16] have made extensive and very successful analyses on the morphological instability of growing epitaxially strained dislocation-free solid films. These analyses were based on the surface diffusion driven by the capillary forces and misfit strains by elaborating various type of wetting potentials associated with the thickness dependent surface specific free energy. In their work, similar to the simulation studies of the stability of epitaxially strained islands by Chiu and Gao,[10] Zhang and Bower,[17] Srolovitz,[14] Krishnamurthy and Srolovitz,[18] Medhekar and Shenoy,[19] Golovin et al.,[20] and Levine et al.[21] elastic strain energy density appears to be additive. Almost without exception, including the work on the equilibrium morphologies by Kukta and Freund,[22] all numerical and analytical studies reported in the literature for the so-called steady state solutions of the nonlinear free moving boundary value problem utilized the periodic boundary conditions, and relied mostly on the instabilities initiated by the white noise or the small amplitude initial perturbations, where the film thickness is smaller than the wavelength of surface variations.

In the present study, we demonstrated that without even imposing any external perturbations on the otherwise smooth surface of droplets, this isochoric composite system (film/substrate) simultaneously evolves towards the stationary state in the absence of the growth mode by creating the SK islands or other proper morphologies depending on the imposed external and internal parameters. Unfortunately, the application of the rigid boundary conditions of any type to the computation domain restricts the natural motions of the triple junction (TJ) that lines between the isolated islands and the substrate, and thus the spontaneous evolution kinetics of the ensemble towards the possible stationary state morphologies are partially hindered. In this work, this restriction on the TJ motion is lifted by employing an irreversible thermodynamic connection obtained by using the internal entropy production (IEP) hypothesis.[23] IEP hypothesis furnishes the temporal velocity of the TJ singularity with the instantaneous values of the contact angle (i.e., one sided dihedral angle) and the wetting parameter, which depends only on the specific surface Helmholtz free energies of the thin film, substrate and the interface between them. In the absence of the growth term, there are at least two important morphological features come into the scenario from this non-equilibrium approach to the free-moving boundary value problem, namely: the *zero-contact angle* for SK islands at the stationary state, and the substantial amount of spreading of the wetting layer associated with the droplet stand beyond the original domain size. During the transient stage, the SK island morphology looks very similar to the topography of the composite solution calculated by Spencer[9] using the "glued-wetting-layer" model. However, in the stationary regime SK island describes strictly monotonic decrease in the profile while approaching to the perfectly flat and highly extended platform with a relatively sharp turn. This plateau corresponds to the wetting layer, which has almost uniform thickness, which is very close to the prescribed thickness of the boundary layer, namely fraction of a nanometer. The transient state profiles look very similar to the ones reported by Tekalign and Spencer[15,16] using the steady state solutions of their non-linear equation with periodic boundary conditions, but without showing any finite thickness for the boundary layer at the turning point (see Fig. 4 in Ref. [15]). Inclusion of the growth term drastically alters the island morphology such that whenever the growth rate exceeds certain limits *the satellite formation* is observed without any excess peak broadening on the central island compared to the height enhancement. On the other hand, lower growth rates cause the replacement of the boundary layer plateau with a bump shape Frank-van der Merwe (FM) type layered structure with relatively large contact angles at the spreading edges. The addition of the growth term accelerates the spontaneous spreading of the wetting layer or platform, which is a very important step for the easy transformation of the VW type clusters into SK islands. At first, VW morphology is observed on those substrates having extremely high film/substrate interfacial energies because of the large misfit ratio. Later, the transformation into SK island morphologies starts by thermal annealing if there is enough relaxation in strain, which initially caused by the misfit dislocation creation at the interface,[24] by establishing direct contact between individually formed SK islands at the initial stage of the continuous wetting layer.

## II. PHYSICAL AND MATHEMATICAL MODELING

A continuum theory based on the microdiscrete formulation of the irreversible thermodynamics of surfaces and interfaces, which was extensively elaborated and applied by Ogurtani[23] and Ogurtani and Oren[25] for multi-component systems has been enlarged by taking into account the film thickness dependent surface Helmholtz free energy (isochoric system) to study the evolution behavior of epitaxial films, especially for the formation of Stranski-Krastanow islands by computer simulations.

### a. The governing equation for the surface drift-diffusion and growth:

The evolution kinetics of surfaces or interfacial layers (simply- or multiply-connected domains) of an *isochoric multi-phase system* may be described in terms of surface normal displacement velocities $\bar{V}_{ord}$ by the following well-posed free moving boundary value problem in 2D space for ordinary points (i.e., the generalized cylindrical surfaces in 3D space) using normalized and scaled parameters and variables, which are indicated with the bar signs. Similarly, the TJ longitudinal velocity $\bar{V}_{Long}$ associated with the natural motion of the droplet-substrate contour line may be given in terms of the wetting parameter $\lambda = [(f_s - f_{ds})/f_{d/s}]$, and the temporal one-sided dihedral or wetting contact angle $\theta_W$ as a dynamical variable. Here $f_s$ is the Helmholtz surface free energy of the substrate, and $f_{ds}$ is the interfacial free energy between the droplet and the substrate, and $f_{d/s}$ is the height dependent surface free energy of the droplet. In the case of Chiu and Gao[10] type model the wetting parameter defined as such that it becomes identically equal to unity $\lambda_W \to 1$. In reality it should take values greater than unity in order to have wetting phenomenon to proceed spontaneously (i.e., natural change). According to our adopted sign convention, the negative values of $\bar{V}_{ord}$ and $\bar{V}_{Long}$ correspond to the local expansion and/or growth of a droplet. Then one writes;

$$\bar{V}_{ord} = \frac{\partial}{\partial \bar{\ell}} \left[ \frac{\partial}{\partial \bar{\ell}} \left( \Delta \bar{f}^o_{dv} - \Sigma(\bar{\sigma}_h)^2 + \bar{f}_{d/s}(\bar{y})\bar{\kappa} + \bar{\omega}(\bar{y}) \right) \right] \\ - \bar{M}_{dv} \left( \Delta \bar{f}^o_{dv} - \Sigma(\bar{\sigma}_h)^2 + \bar{f}_{d/s}(\bar{y})\bar{\kappa} + \bar{\omega}(\bar{y}) \right)$$  (Ordinary points)   (1)

and

$$\bar{V}_{Long} = -\bar{M}_{Long} \bar{\Omega}^{-1} \{\lambda - \cos(\theta_W)\} \quad \forall \ \lambda \geq 1$$  (Triple junction contour)   (2)

Where $\lambda$ is the wetting constant as defined previously, and $\theta_W$ is the dihedral or contact angle, which varies in the range of $\theta_W \subset \left(0^o - \pi\right)$, zero degree corresponds to full wetting. The variables in Eq. (1) and (2) are described as follows: $\bar{\Omega} \cong 10^{-3}$ is the normalized atomic volume in the particle representation by assuming tentatively that the scaling length is in the range of 10 atomic spacing (for the more details see Ref. [23] and Ref. [25]).[26] $\bar{M}_{Long}$ is the ratio of the mobility of the TJ, $\hat{M}_{Long}$, to the surface mobility, $\hat{M}_d$. Similarly, $\bar{M}_{dv}$ is the normalized growth mobility, which in general may depend on the temperature and the surface stress.[27] $\bar{\kappa}$ is the normalized local curvature and is taken to be positive for a concave solid surface (troughs), and the positive direction of the surface displacement and the surface normal vector $\hat{n}$ are assumed to be towards the bulk (i.e., droplet) phase, and implies the local shrinkage or evaporation processes. In the governing equation, Eq. (1), the normalized hoop stress is denoted by $\bar{\sigma}_h \equiv Tr\bar{\sigma}$, where the dimensionless stress intensity parameter $\Sigma$ corresponds to the intensity of the elastic strain energy density (ESED) contribution on the stress-driven surface drift diffusion. The misfit strain $\varepsilon_o$ at the film/substrate interface is introduced as a *Dirichlet boundary condition* by specifying the displacement vector in 2D space as $\tilde{u} \to \hat{i}\varepsilon_o x$ (i.e., in 3D pseudo-space $\tilde{u} \to \left[\hat{i}\varepsilon_o x, \hat{k}\varepsilon_o z\right]$), and taking the droplet center at the film/substrate interface as the origin of the coordinate system to avoid shifting. Similarly, the stress used for the normalization procedure is chosen as the biaxial stress $\sigma_o = E_d \varepsilon_o / \left(1 - v_d\right)$, where, $E_d$ and $v_d$ are, respectively, Young modulus and Poisson's ratio of the droplet shape film, and $\varepsilon_o$ is misfit strain at the film/substrate interface. This choice is very convenient for the indirect boundary elements method (IBEM) solution of the plain strain isotropic elasticity problems,[28] where one takes $E_d \to 1; \varepsilon_o \to 1$ as the initial scaling data. Then, only the actual value of the Poisson's ratio of the film has to be supplied for the computation of the normalized stress distribution. The rest $\{E_d; \varepsilon_o\}$ is embedded in the definition of $\Sigma$. Here, we assumed that the surface Helmholtz free energy density $f_{d/s}(y)$ for an isochoric system depends on the local distance $y$ between the surface layer and the substrate, and the special form of which will be presented later in this section. $\bar{\omega}(y)$ is the normalized wetting potential, which is given by $\omega(y) = \Omega_d n_y df_{d/s}/dy$ in particle representation, where $n_y = -\hat{n}.\hat{j}$ is the projection of the surface normal along the y-axis, which is taken as perpendicular to the substrate. The second group of terms in governing equation Eq. (1) is related to the growth or phase transformation (*condensation or evaporation*) kinetics. In the above expression, $\bar{\ell}$ is the curvilinear coordinate along the surface (arc length) in 2D space scaled with respect to $\ell_o$. Where, $\ell_o$ is the arbitrary length scale, which may be selected as the peak height of the droplet or the ratio of the surface Helmholtz free energy of the film in the bulk to the elastic strain energy density[10,16] such as $\ell^* = f_d/w_o$. Here, $w_o = (1-\nu_d^2)\sigma_o^2/2E_d$ denotes ESED, which is associated with the nominal biaxial misfit stress taking the third dimension into account. If one takes $\ell^*$ as a length scaling parameter then one should have the following replacement: $\Sigma \leftarrow 1$, since according to our definition $\Sigma \equiv \ell_o/\ell^*$. In the present paper, otherwise it is stated, the initial peak height of the droplet $h_p$ is chosen as the natural scaling length, namely; $\ell_o = h_p$. The film thickness $h_o$ is defined as the *integrated film thickness,* and it may be given by $\bar{h}_o = 2\bar{h}_p / \pi \to 0.637$ for the scaled halve wave length Cosine-shape flat droplets, where $\bar{h}_p \to 1$. $\Delta \hat{f}_{dv}^o(T) = \left( \hat{f}_v^o - \hat{f}_d^o \right)$ represents the thermal part of the Helmholtz free energy of transformation for a flat interface assuming that the isothermal processes is taking place in an isochoric system. The positive value corresponds to condensation of the vapor phase or for the growth of the droplet. $\hat{f}_v^o$ and $\hat{f}_d^o$ are the volumetric Helmholtz free energy densities, respectively, for the realistic vapor and bulk droplet phases.

In the present enlarged formulation of the problem, as we did in Ref. [25], we scaled the time and space variables $\{t, \ell\}$ in the following fashion: first of all, $\hat{M}_d$ an atomic mobility associated with the mass flow at the surface layer is defined, and then a normalized time scale is introduced by $\tau_o = \ell_o^4 / \left( \Omega^2 \hat{M}_d f_d \right)$. The bar signs over the letters indicate the following scaled and normalized quantities:

$$\bar{t} = t/\tau_o, \ \bar{\ell} = \ell/\ell_o, \ \bar{\kappa} = \kappa \, \ell_o, \ \bar{L} = L/\ell_o, \ \Delta \bar{f}_{dv}^o = \frac{\Delta \hat{f}_{dv}^o}{f_d} \ell_o, \ \bar{\sigma}_h = \frac{\sigma_h}{\sigma_o} \qquad (3)$$

$$w_o = \frac{\left(1 - v_d^2\right)}{2 E_d} \sigma_o^2, \ \sigma_o = \frac{E_d}{(1 - v_d)} \varepsilon_o, \ \Sigma = \frac{\left(1 - v_d^2\right) \ell_o}{2 E_d f_d} (\sigma_o)^2 \equiv \frac{\ell_o}{\ell^*} \qquad (4)$$

and

$$\bar{\omega}(\bar{y}) = \frac{1}{\sqrt{1 + \bar{y}_x^2}} \frac{(f_s - f_d)}{\pi f_d} \frac{\bar{\delta}}{\bar{\delta}^2 + \bar{y}^2} \qquad (5)$$

and,

$$\bar{f}_{d/s}(y) = \frac{(f_d + f_s)}{2f_d} + \left(\frac{(f_d - f_s)}{f_d}\right)\frac{1}{\pi}\arctan(y/\delta) \quad \forall \quad \left(\lambda \equiv \frac{f_s - f_{ds}}{f_{d/s}} \rightarrow \frac{2(f_s - f_{ds})}{f_s + f_d}\right) \geq 1 \quad (6)$$

Here, we adapted a transition-layer model as advocated by Spencer,[9] but reserving the case for the description of the wetting constant $\lambda$ since Spencer[9] and his coworkers[15] assumed that the interfacial free energy between the droplet and the substrate, $f_{ds}$, is negligible. Similar to the definitions of $f_s$ and $f_{ds}$, $f_d$ is the surface energy of the droplet in the bulk form. According to the functional relationship given in Eq. (6) for the boundary layer model, the film specific Helmholtz surface free energy undergoes a rapid transition from film, $f_d$, to substrate, $f_s$, values over a length scale denoted as $\delta$. That means when the film thickness becomes equal to zero (i.e., at the TJ) the droplet surface free energy should be given by $f_{d/s} \rightarrow (f_s + f_d)/2$. This may create some trouble at the thermodynamic equilibrium, because of the fact that the TJ configuration for the full wetting is defined by $\lambda = 1$ for the reversible changes. This condition requires existence of the following equality between the surface free energies, $f_{ds} = 1/2(f_s - f_d)$ if one adopts the Spencer's model. Then one would have zero contact angle at the TJ without pushing the wetting potential to nil. Similarly, the equilibrium wetting or contact angle, which is given by $\theta_{Eq.} = \arccos\{(f_s - f_{ds})/f_{d/s}\}$ becomes *undefined* at the droplet/substrate TJ unless $\lambda = 1$ if one assumes that the relationship Eq. (6) is still holds along the contour contact line. This is an important dilemma of their model, since this assumption results a zero wetting potential as defined also by Spencer.[9] The model proposed by Chiu and Gao[10] partially removes this awkward situation but brings restrictions to the TJ motion for the quasi-equilibrium case, $\lambda = 1$. But this is not a necessary condition for the non-equilibrium stationary states. In general, the wetting parameter should be defined by $-1 \leq \lambda_W = [(f_s - f_{ds})/f_d] \leq 1$, which covers the complete range of thermodynamic equilibrium wetting or sticking phenomenon, namely; starting from the complete wetting configuration (i.e., $\theta_W = 0$) to the another extreme case of *point contact* (i.e., $\theta_W = \pi$), which may be described as a rigid ball sitting at the top of the substrate, at the TJ. In the present computer simulations similar to Spencer[9] and his coworkers,[15] we assumed that $f_{ds} \cong 0$ for the wetting potential, which is very plausible for the coherent boundaries such as the interface between epitaxially grown film and the substrate.

In the present study, the generalized mobility, $\bar{M}_{dv}$, associated with the interfacial displacement reaction (adsorption or desorption) is assumed to be independent of the orientation of the interfacial layer in crystalline solids. As we already mentioned, This generalized mobility is normalized with respect to the minimum value of the mobility of the surface drift-diffusion denoted by $\hat{M}_d$. They are given by: $\bar{M}_{dv} = (\hat{M}_{dv} \ell_o^2)/\hat{M}_d$ and, $\hat{M}_d = (\tilde{D}_d h_d / \bar{\Omega} kT)$ where, $\bar{\Omega}$ is the mean atomic volume of chemical species in the surface layer and $\tilde{D}_d$ is the isotropic part (i.e., the minimum value in the case of anisotropy) of the surface diffusion coefficient.

**b. Linear instability analysis with the height dependent surface free energy:**

We presented the general treatment of the unified linear instability analysis (ULISA) of the governing equation, Eq. (1), in Ref. [27] by taking any possible diffusivity and surface free energy anisotropies into account.[27] However, in that theory, we did not consider the wetting effect, which we are going to launch now in this paper. The wetting effect may be included by linearizing the wetting potential at the position of the original flat film surface denoted as $h_o$ (i.e., defined as the integrated thickness of the droplet of any shape) by using the local tangent line for the linear extrapolation. This is a similar approach, introduced by Pang and Huang,[29] and gives the growth rate constant in terms of the critical thickness, $h_c$, normalized wave number, $\underline{k} = k\ell^*$, and the time scale $\tau_o^* = \ell^{*4} / \left(\Omega_d^2 \hat{M}_d f_d\right)$. Then, one obtains the following expression for the normalized growth rate constant:

$$\underline{\Gamma} \simeq -\left\{-4\underline{k} + \underline{k}^2 + 4\left[\frac{h_o}{h_c}\right]^{-3}\right\}\underline{k}^2 \; ; \quad h_c = \ell^*\left(\frac{(f_s - f_d)}{2\pi f_d \ell^*}\delta\right)^{1/3} \; ; \quad \ell^* = \frac{f_d}{w_o} \; ; \tag{7}$$

There is a minor difference in our expression and the one presented by Pang and Huang[29] (see Eq. (35) in Ref. [29]), which arises from the definition of the length scale employed by those authors and results a factor of 4 smaller value than ours.

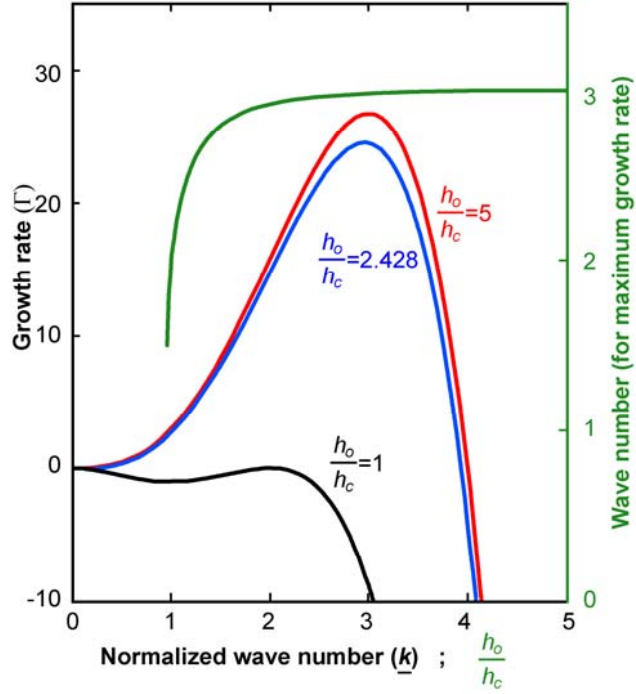

**FIG. 1** (Color online) Linear analysis of the wetting effect: The growth rate constant, $\Gamma$, is given on the left *y*-axis, with respect to the normalized wave number, $\underline{k} = k\ell^*$, for various normalized film thicknesses $\{h_o/h_c\}$. On the right *y*-axis, the wave number for the maximum growth rate is also presented with respect to the normalized film thickness $\{h_o/h_c\}$ that is scaled with respect to the critical film thickness denoted as $h_c$.

The formulation of Pang and Huang[29] has another problem in the definition of ESED, where a factor of two is missing in the dominator of their Eq. (11) in Ref. [29]. In Eq. (7), $\delta$ is a characteristic length typically encountered in a boundary layer model, and determines the size of the transition region. Since arctan function is a relatively slow decaying function, one needs $\delta/\ell^* \sim 0.01 - 0.005$ to be smaller than the effective wetting layer thickness by a factor of 50 as suggested by Tekalign and Spencer,[15] so that the wetting layer transition should occurs at about one atomic spacing. In Fig. 1, the plots for the normalized growth rate constant versus normalized wave number are presented for various values of film thickness versus critical thickness ratios $[h_o/h_c]$. The *instability band width* for the normalized wave number may be easily calculated form Eq. (7) as $\underline{k}_{1,2} = 2\left(1 \pm \sqrt{1-(h_o/h_c)^{-3}}\right)$. The maximum growth rate occurs at the normalized wave number denoted by $\underline{k} = 3$, when $h_o/h_c \geq 5$, which is identical to the one reported by Chiu and Gao,[10] even though they were using a different functional representation for the surface free energy but for the same formula for the ESED. According to the present theory, the upper limit for the normalized wave number for the instability range is given by $\underline{k}_2^{\max} = 4$.

c. **Implementation of the IBEM numerical method:**

The detailed description of the indirect boundary elements method[30] (IBEM), and its implementation are presented very recently in two comprehensive papers by Ogurtani and Akyildiz[31,32] in connection with the void dynamics in metallic interconnects. In both papers, the void dynamics were driven by the surface drift-diffusion induced by the inhomogeneous thermal stress fields in collaboration with the electromigration as the main driving forces. In this study, we utilized the simplest implementation of the IBEM that employs the straight constant line elements in the evaluation of the hoop stress at the free surface of the droplet, as well as along the interface between droplet and the substrate. In fact, it is also possible to generate the complete stress distribution field in the interior region of the sample as a byproduct. Here, Neumann (i.e., traction free boundary condition) and Dirichlet boundary conditions (i.e., prescribed displacements) are utilized, respectively, along the free surface of the droplet and at the interface between droplet and the substrate. Therefore, we have implicitly assumed that the substrate is rigid, and the displacement is supplied as a Dirichlet boundary condition along the interface, which is calculated from the misfit strain, $\varepsilon_o$, by $u_x(0) = \varepsilon_o x$. This implementation, adopted by the present author, guarantees the surface smoothness conditions for the validity of the governing Fredholm integral equation of the second kind at the corners and edges, which may be generated artificially during the numerical procedure. The explicit Euler's method combined with the adapted time step auto-control mechanism is employed in connection with Gear's stiff stable second-order time integration scheme[30] with the initial time step selected in the range of $\left(10^{-8} - 10^{-9}\right)$ in the normalized time domain. This so-called adapted time step procedure combined with the self-recovery effect of the capillary terms guarantees the long-time numerical stability and accuracy of the explicit algorithm even after performing $2^{45} - 2^{50}$ steps, which is clearly demonstrated in our recent work on the grain boundary grooving and cathode drifting in the presence of electromigration forces.[32] The network remeshing is continuously applied using the criteria advocated by Pan and Cocks,[33] and the curvature and normal line vector are evaluated at each node using a discrete geometric relationship in connection with the fundamental definitions of the radius of curvature.

## III. RESULTS AND DISCUSSION

In our computer simulation studies, it is assumed that the thin film on top of the substrate is represented by a flat crystalline droplet (i.e., bump), which may be described by a symmetrically disposed, halve-wave length *Cosine-function* having a wave length and a height (i.e., amplitude) denoted by $2L$ and $h_p$, respectively. The droplet *aspect ratio* may be defined by: $\beta = L/h_p$, which prescribes a finite contact angle $\theta = \arctan(\pi/\beta)$ between film and the substrate at the onset of the simulation run. Therefore, in the normalized and scaled time-length space, the initial shape of a droplet is uniquely described by one single parameter, namely the aspect ratio $\beta$, since $\bar{h}_p = 1$ according the scheme adopted in this study. Similarly, a close inspection of the normalized governing equation without the growth term shows that there is only one more additional parameter left for the complete predetermination of the morphological evolution process as an initial data, which is the ESED parameter denoted by $\Sigma = \ell_o / \ell^* \rightarrow h_p / \ell^*$. In real space, the size of the droplet may be described by $h_p = \ell_o$ for a given value of the aspect ratio, but now it is solely determined by $\Sigma$ keeping the shape invariant (i.e, *zooming*) due to the fact that the characteristic length $\ell^*$ is an internal variable for the isochoric systems, and it depends only on the material properties of the film and the substrate including the misfit strain. Therefore, this unitless parameter $\Sigma$ completely dictates the possible *size effects* of the droplet on the evolution process in real space; where one has: $h_p \leftarrow \ell_o = \Sigma \ell^*$. Hence, in the absence of the growth term, the aspect ratio $\beta$ (i.e., shape) and the strain energy density parameter $\Sigma$ (i.e., size) are two basic numbers capable to dictate the topographic features of the final stationary states as will be demonstrated later in this study.

    Physically, the droplet is attached to the substrate with a coherent interface, and the top surface is subjected to the surface drift diffusion, and it is exposed to a vapor environment, whose pressure may be neglected. Since we are performing 2D simulations

(equivalent to parallel ridges or quantum wires in three dimensions), no variation of the interface profile and the displacement fields in the film and substrate occurs in the direction (i.e., $\hat{z}$ axis) perpendicular to the plane of the schematics in Fig. 2(a) (i.e., plane strain condition). Similarly, to simplify the numerical computations we assumed that the film/substrate interface is flat and the substrate is stiff. These assumptions guarantees that the initial displacement along the interface associated with the misfit strain $\varepsilon_o$ stays constant during the evolution process (i.e., Dirichlet boundary condition). In the simulation studies, scaled and normalized variables which were thoroughly described in our previous publication[32] are utilized, and the length scaling is performed with respect to the droplet initial peak height designated as $\ell_o = h_p$.

During the evolution process, the shape of the surface profile changes continuously. Thus, one has to utilize the power dissipation concept to calculate the global strain energy change of the droplet by taking the time derivative of the bulk Helmholtz free energy variation for an infinitesimal displacement of the surface layer along the surface normal designated as $\delta\eta$, which is given by the relationship $\delta W(\eta) = -\int_{Surf.} (w_d)\delta\eta d\ell$ for the traction free surfaces. According to our adopted definition of the surface normal, which is directed towards the solid phase, the shrinkage and the expansion of the solid phase is respectively corresponds the inequalities $\delta\eta = -\delta h > 0$ and $\delta\eta = -\delta h < 0$. These results are in complete accord with findings by Rice and Drucker[34] and Gao[35] for the traction free surfaces. In a more general case, Eshelby[36] found similar expression with an additional term related to *the energy-momentum tensor* for the bimaterial interfaces, which may carry non-vanishing tractions.

Here it has been presumed that the interface between film and substrate is immobile. Then one reads the following expression in the normalized time-length space, where '$n$' designates the total number of nodes along the traction free surface, and $\bar{L}(\bar{t})$ is the instantaneous length of the free surface contour.

$$\bar{P}(\bar{t}) = - \int_{Surf.} \left(w^b\right) \dot{\eta} d\ell \Rightarrow -w_o \bar{L}(\bar{t}) \sum_{j=0}^{n-1} \frac{\left(\bar{\sigma}_j(\bar{t})\right)^2}{n} \dot{\eta}_j(\bar{t}). \qquad (8)$$

In Eq. (8), we did not include the contribution associated with the time variations in the strain energy density distribution evaluated at the free surface. Because, the numerical calculations showed that it is three orders of magnitude smaller than the one presented above with the same trend, i.e., both having a negative sign. Subsequently, the cumulative change $\Delta \bar{F}_d$ in the bulk Helmholtz free energy, which is equal to the total elastic strain energy for the isothermal changes, during the evolution process may be calculated as a function of the discrete normalized time $\bar{t}_m$ by using a simple integration (i.e., summation) procedure applied to above expression. This procedure yields:

$$\Delta \bar{F}_d \left(\bar{t}_m\right) \equiv \Delta W(\bar{t}_m) = \int_0^{\bar{t}_m} d\bar{t}\, \bar{P}(\bar{t}) \Rightarrow \bar{t}_m \sum_{k=0}^{k=m} \bar{P}(\bar{t}_k)/m. \qquad (9)$$

Similarly, one may also write the Helmholtz surface free energy change $\Delta F_s$ associated with the free surface contour enlargement with respect to the initial configuration, which may be given for a prescribed time, $\bar{t}$, as;

$$\Delta \bar{F}_s(\bar{t}) = f_d\left[\bar{L}(\bar{t}) - \bar{L}(0)\right] = w_o \ell^*\left[\bar{L}(\bar{t}) - \bar{L}(0)\right] \tag{10}$$

In this paper, the bulk and surface Helmholtz free energy plots are normalized with respect to the nominal strain energy density $w_o \rightarrow (1+\nu)E\varepsilon_o^2/2$ to compare them properly even in the normalized time-length space. For the future application in Ge/Si (100) system, the nominal elastic strain energy density may be given by $w_o \cong 1.58 \times 10^8 \ N/M^2$. Our computer simulation shows that one always observes the fulfillment of the following inequalities during the spontaneous evolution processes, $\Delta \bar{F}_d(\bar{t}) < 0$ and $\Delta \bar{F}_s(\bar{t}) > 0$. Even though their straight summation in normalized space may not be negative, one still expects for the natural isothermal processes occurring in the isochoric systems, the inequality $\Delta F_d(\bar{t}) + \Delta F_s(\bar{t}) < 0$ should be satisfied in the real time and length space. In fact, this requirement is also found to be satisfied in all the computer simulation experiments presented in this paper. Since the numerical calculations are carried out in normalized and scaled space, the following connections between the normalized and the real Helmholtz free energies associated with the elastic strain and the surface free energy contributions become very important. One may obtain these connections using the dimensional analysis as: $\Delta F_d = \ell_o^2 \Delta \bar{F}_d$ and $\Delta F_s = \ell_o \Delta \bar{F}_s$. Finally, these expression may be converted into following forms:

$$\Delta F_d(t) = \Sigma^2 \ell^{*2} \Delta \bar{F}_d(\bar{t}) \quad \text{and} \quad \Delta F_s(t) = \Sigma \ell^* \Delta \bar{F}_s(\bar{t}) \tag{11}$$

### a. Morphological evolution of droplet without growth term:

In this section, we will present the results obtained from a set of special computer experiments done on the specimens having large aspect ratios (i.e., in the range of $\beta = (10-28)$), and subjected to the misfit strain at the interface between the thin film and the stiff substrate. Here, the aspect ratio is defined as the ratio of the droplet base length-to-peak height. The surface of the droplet film initially is represented by a symmetrically disposed halve-wave length Cosine-curve as illustrated in Fig. 2(a), which has a normalized base length of $\beta = \bar{L} = 28$, and an amplitude of $\bar{h}_p = 1.0$ as compared to the integrated film thickness, which is given by $\bar{h}_o = 2\bar{h}_p / \pi \rightarrow 0.637$. Although, we employed a large number of different elastic strain energy density parameters (ESED) in our experiments, we will discuss only a few that represents different parts of the spectrum of morphologies; (i.e., $\Sigma = 0.075, 0.175, \|0.250\|, 0.300, 0.350, 0.400, \|0.413\|, 0.425, 0.450$). The double vertical lines, $\| \|$, roughly indicate the transient states found. The lowest ESED value presented here is $\Sigma = 0.075$, which destabilize the initial droplet configuration by activating the TJ towards the Frank-van der Merwe layer structure by spreading over the substrate surface before switching to the island formation. Here we observed not only the singlet but also the island doublets (i.e., twins, etc) separated by the transient morphologies. During our simulations, besides film morphologies, we also monitored the kinetics of the peak height development, the displacement of the TJ singularity during wetting layer extension, and the strain energy release during the evolution process. In order to illustrate the actual physical size of the islands, we consider the following parameters,[37] which are representative of Ge film grown epitaxially on a stiff silicon substrate.[7,37] Namely: $\varepsilon_o = -0.042$, $E_{Ge} = 103$ GPa, $\nu_{Ge} = 0.273$, $f_{Ge} = 1.927 Jm^{-2}$, and $f_{Si} = 2.513 Jm^{-2}$. These numbers imply a characteristic length of $\ell^* = 12.11$ nm, which may be used to calculate the heights and the base lengths of the droplets that are corresponding to the range of the strain energy intensity parameters for a given aspect ratio (i.e., $\beta = 28$), namely; for the singlet islands one has: $\{\Sigma : 0.25 - 0.40\} \Rightarrow \{h_o = 1.91\ nm - 3.08\ nm\} \cap \{L : 84\ nm - 134\ nm\}$.

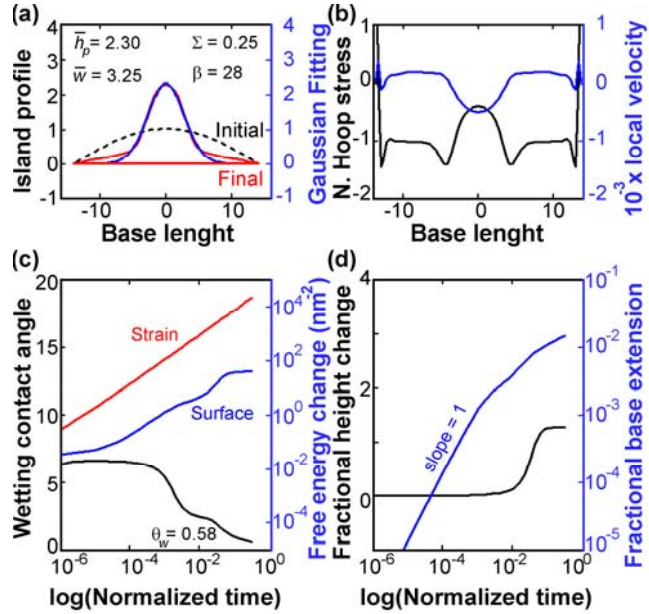

**FIG. 2. (Color online)** (a) The initial and the final stationary island profile at the transient stage just before the onset of the SK islands formation regime. The final profile is well represented by a Gaussian bell-shape curve having following parameters: $\bar{h}_p = 2.30$, $\bar{w} = 3.25$, which corresponds to the peak height-to-peak width ratio of $\xi = 0.354$. (b) Instantaneous velocity and the hoop stress distributions along the final droplet profile. (c) Evolution of the contact angle is shown on the left *y*-axis. On the right *y*-axis, the strain energy and surface free energy changes are given for Ge/Si(100) system, and scaled by $nm^2 \rightarrow 10^{-18}$. (d) Time evolution of peak height and TJ displacement.

Simulation Data: $\Sigma = 0.25$, $\bar{h}_p = 1$, $\beta=28$, $\nu=0.273$, $\bar{M}_{TJ} = 2$, $\lambda=1$, $\bar{\delta} = 0.005$, $f_s = 1.2$ and $f_d = 1$.

The results of a computer simulation, which is done on a hypothetical sample by assigning a critical value for the elastic strain energy density parameter (ESED) such as $\Sigma = 0.25$ are presented in Fig 2. Fig. 2(a) shows development of a premature or transient island profile without having any indication of the wetting layer formation even after $2^{46} \simeq 7 \times 10^{13}$ runs. This profile, which was obtained by performing numerous experiments in the vicinity of the stability-instability turn-over point for the linearized systems, corresponds to the transient stage between the SK islands and the FM type layers structures. This final profile as demonstrated in Fig. 2(a) may be described by a Gaussian curve (i.e., second degree) given by $G(x;\bar{h}_p,\bar{w}) = \bar{h}_p \cdot \exp(-ln(2)x^2/\bar{w}^2)$, having a halve-width of $\bar{w} = 3.25$, and a peak height of $\bar{h}_p = 2.30$, in normalized space. These two value, corresponds to the peak height to peak width ratio of $\xi = 0.354$. According to the Prigogine[38] description, this is a genuine *stationary non-equilibrium state* since even though the height of droplet reached a plateau region (Fig. 2(d)), the TJ contour line is still active with a temporal wetting angle of $\theta_W \approx 0.58^o$ (Fig. 2(c)). This TJ activity is the main indicator that the system is in the non-equilibrium state. To reveal the real physical system parameters, we employed the data given above for the Ge/Si(100) system to the normalized and scaled parameters and obtained $h_p \simeq 6.9$ *nm* for the peak height, $2W = 18.98$ *nm* for the peak width of and $L \equiv \lambda \approx 86.5$ *nm* for the base (or the wave length that describes the spacing between islands) length with the help of Fig. 2(b).

These values are in the range of numbers reported by Kukta and Freund,[22] who were defining the base of the island as its width, which may create some confusion if there is no sharp turning point at the corners that separate island from the wetting layer (see Fig. 5 in Ref. [22]). As seen in Fig. 2(d), the peak height showed logarithmic time dependence during the intermediate regime before the onset of the plateau region, namely; $\bar{h}_p(\bar{t}) \cong 2\log(\bar{t}) + 3.6$. In Fig. 2(c), the *negative* cumulative strain energy release, $-(\Delta F_d / w_o) nm^2$, and the surface free energy variation $(\Delta F_s / w_o) nm^2$, both scaled with respect to $w_o$, are plotted using the connections given in Eq. (11) for Ge/Si(100). This plot shows almost perfectly linear decrease for the cumulative strain energy release with time compared to the surface free energy variation that indicates a leveling off in the early stages of the development followed by a positive change due to the surface layer extension during the evolution process. The free energies are plotted by considering the critical length of Ge/Si (100) system, which is about 12.11 *nm* for the present case. At the end of the test run the total strain energy release is calculated to be about $\Delta F_d \cong -3.203 \times 10^{-5} J$, which is very large compared to the total surface energy gain that amounts to $\Delta F_s \cong 3.36 \times 10^{-9} J$. This figure also shows that the global Helmholtz free energy is negative all the way through the natural change as one should expect from the thermodynamic considerations.

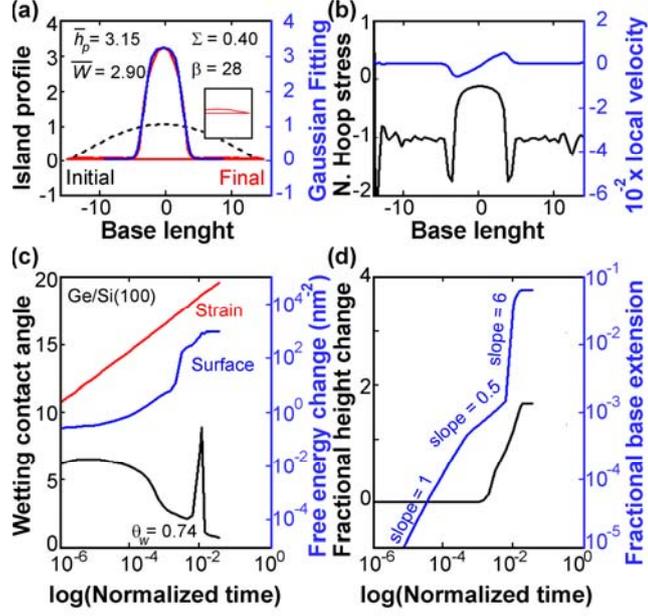

**FIG. 3. (Color online)** (a) Spontaneous formation of the SK island with an almost perfect flat wetting layer from a single crystal droplet on a stiff substrate by the isotropic surface drift diffusion driven by the combined actions of the misfit strain (isochoric) and the capillary forces. The inset details the structure of the wetting layer at the TJ. Gaussian bell-shape curve representing the final profile has the following parameters: $\bar{h}_p = 3.15$, $\bar{w} = 2.90$ and thus the peak height-to-peak width ratio of $\xi = 0.543$. (b) Instantaneous velocity and the hoop stress distributions along the final droplet profile. (c) Evolution of the contact angle is shown on the left *y*-axis. On the right *y*-axis, the strain energy and surface free energy changes are given for Ge/Si(100) system. (d) Time evolution of peak height and TJ displacement. Simulation Data: $\Sigma = 0.40$, $\bar{h}_p = 1$, $\beta = 28$, $\nu = 0.273$, $\bar{M}_{TJ} = 2$, $\lambda = 1$, $\bar{\delta} = 0.005$, $f_s = 1.2$ and $f_d = 1$.

In Fig. 3, a typical morphological evolution behavior of the SK island is presented in terms of the final droplet profile, the peak height, the base extension, and the TJ contact angle with respect to the normalized logarithmic scale. In this experiment, we utilized an elastic strain energy density parameter of $\Sigma = 0.40$ that was picked out from the upper edge of the stable singlet SK island formation range $\{\Sigma: 0.30, 0.35, 0.40\}$. The SK profile reported in this figure shows a very thin simultaneously-formed wetting layer having a normalized thickness of $\Delta \bar{h} \simeq 0.026$. This wetting layer thickness is about a factor of 5 greater than the adopted boundary layer thickness in our computer simulations, which enters as $\bar{\delta} = 0.005$ into the wetting potential presented in Eq. (1). In real space, the wetting layer thickness for the Ge/Si(100) system may be computed as follows: $\Delta h \simeq 0.026 \ell_o \rightarrow 0.026 \Sigma \ell^* \simeq 0.12\ nm$, which may be easily improved by taking the boundary layer thickness 5 times smaller than the desired effective wetting layer thickness,[15] namely that is about one atomic spacing, $0.6\ nm$. That means one should rather take $\bar{\delta} \rightarrow 0.025$.

The insert in Fig. 3(a) demonstrates the structure of the wetting layer at the TJ contour line, which has a temporal contact angle of $\theta \simeq 0.74^o$ instead of zero degree, which indicates that the TJ is still active. A close inspection of Fig. 3(d) shows that the TJ displacement motion indicates three different time exponent stages, $\bar{L}(\bar{t}) = A\bar{t}^n$ where $n = 1;\ \frac{1}{2};\ 6$, before it enters to the plateau region. Similarly, the peak height shows a logarithmic time dependence during the intermediate regime before the onset of the plateau region, namely; $\bar{h}_p(\bar{t}) \sim 2\log(\bar{t}) + 5.6$. Using the physicochemical data given for Ge/Si(100) system, one may calculate the critical film thickness from Eq. (7), as: $h_c^{Ge} = 0.56\ nm$, and the integrated thickness of the droplet as: $h_o^{Ge} = (2/\pi)\Sigma \ell^* \approx 3.08\ nm$. The critical parameter, which is given by $h_o^{Ge} / h_c^{Ge} \approx 5.546 \geq 5$ is in the range where the wetting parameter does not play any role as may be seen from Fig. 1. The normalized wave number $\underline{k} \equiv k\ell^*$, which corresponds to the maximum growth rate constant, may be calculated from the expression $\underline{k}_{max} = \left\{ 3 + \sqrt{9 - 8(h_c/h_o)^{-3}} \right\}$, which yields $\underline{k}_{max} \approx 2.996$. This result is very close to the theoretical value of 3. The perturbation wave length for the maximum growth rate constant now becomes about $\lambda_{max} = 25.4\ nm$. This figure is about a factor of five smaller than the domain length of $L = 28\Sigma\ell^* \rightarrow 135.7\ nm$. According to the linear instability theory the system should be completely in the instability regime, therefore no stationary non-equilibrium state SK island formation would be possible. This is completely contrary to the findings demonstrated in this work, which implies that for the large amplitudes as well as for the certain initial configurations such as the flat droplets the linear instability theory is not reliable in predicting evolution behavior of the system.

In Fig. 3(c), the cumulative strain energy change, $-(\Delta F_d / w_o) nm^2$, as well as the increase in the surface free energy, $(\Delta F_s / w_o) nm^2$, of the droplet due to the island formation are presented. This figure clearly shows that there is a large increase in the surface free energy due to the island formation compared to Fig. 2(c) because of a factor of two peak height enhancement during the evolution process. Even though the surface free energy levels off after reaching the stationary non-equilibrium state, still the strain energy release continuous to increase due to the readjustment of the system through the TJ activities.

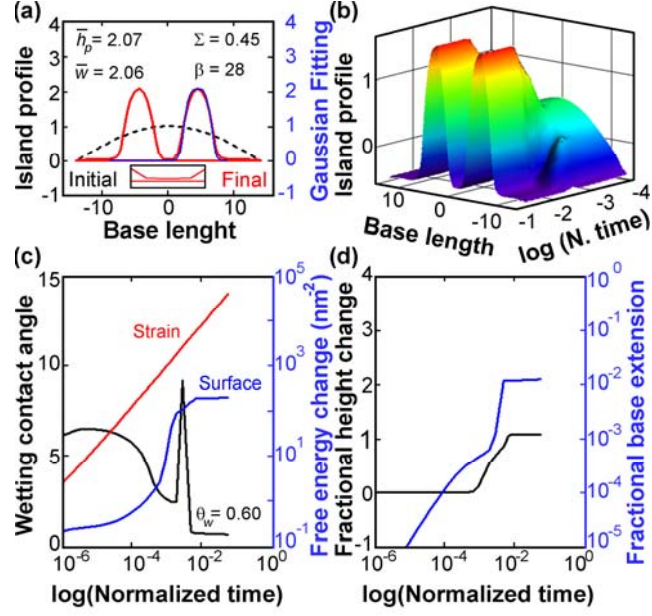

**FIG. 4. (Color online)** (a) Spontaneous formation of the SK doublets with an almost perfect flat wetting layer from a single crystal droplet. The inset details the structure of the wetting layer between the individual peaks. The doublets has fourth degree Gaussian bell-shape curve having the following parameters: $\bar{h}_p = 2.07$, $\bar{w} = 2.06$ and thus the peak height-to-peak width ratio of $\xi = 0.502$. (b) 3D time evolution of island profile. (c) Evolution of the contact angle is shown on the left y-axis. On the right y-axis, the strain energy and surface free energy changes are given for Ge/Si(100) system. (d) Time evolution of peak height and TJ displacement. Simulation Data: $\Sigma = 0.45$, $\bar{h}_p = 1$, $\beta = 28$, $\nu = 0.273$, $\bar{M}_{TJ} = 2$, $\lambda = 1$, $\bar{\delta} = 0.005$, $f_s = 1.2$ and $f_d = 1$.

In Fig. 4(a and b), we illustrate a fully developed SK doublet at the stationary state separated by a thin wetting layer having a thickness of $\Delta \bar{h} \simeq 0.0314 \rightarrow 0.17\ nm$. The wetting layer thickness between the peaks, and the peak tails are found to be almost same. In this case,, we utilized an ESED parameter of $\Sigma = 0.45$, which is selected from a range $\{\Sigma : \|4.125\|; 4.25; 0.45..\}$, where the doublet formation appears to be the stationary state instead of singlets. Above this range not only the multiples but also the Volmer-Weber type island formation may seen depending on the ESED value, which should be further investigated. The extended plateau in the TJ wetting angle plot in Fig. 4(c) indicates that at the stationary state equilibrium contact angle may not be necessarily realized, which should be otherwise zero degree. These doublet peaks may be represented by the fourth degrees Gaussian type function $G(x;\bar{h}_p,\bar{w}) = \bar{h}_p.\exp\left(-ln(2)x^4/\bar{w}^4\right)$, where the peak height and the halve width found to be $\bar{h}_{max} = 2.07 \rightarrow 11.18\ nm$ and $\bar{w} = 2.06 \rightarrow 11.12\ nm$ respectively.

There is a strange peak on the wetting angle plot in Fig. 4(c), and the same phenomenon is also occurred in the formation of the singlet without the sign fluctuation in the global Helmholtz free energy. This event is strongly correlated with the TJ motion as may seen from Fig. 4(d), which shows drastic enhancement in the displacement velocity just at the onset of the stationary non-equilibrium regime.

In Fig. 5, we present a new set of computer simulation studies utilizing an aspect ratio of $\beta = 10$ which is 2.8 times smaller than the first set reported above. As we expected, this modification pushed the onset of the SK island formation threshold described by the ESED parameter to higher values of $\Sigma \rightarrow 0.40$. This is a factor of 1.6 enhancement compared to the case presented in Fig. 2(a). As seen in Fig. 5, we have a bell shape profile extended all over the computation domain without the existence of any wetting layer. This is very typical for this transient regime as observed previously. Figure 5(a and b) shows that there is only a transformation of the Cosine-shape droplet into the second degree Gaussian shape profile with a minor increase in height and a very small stretching of the base line or the computation domain due to TJ motion. Fig. 5(c) indicates that the wetting contact angle reached a value of $\theta_W \cong 1.52^o$, showing some sort of trend towards the plateau behavior. The most interesting event observed here is the sign of the global Helmholtz free energy change during the evolution process: In general it is negative indicating that the decrease in the strain energy is greater than the increase in the surface free energy of the system. However, only in one narrow region, one observes a sign inversion, which indicates the dynamical nature of the simulation experiment due to TJ displacement motion and may be interpreted as this abrupt change is unnatural. Nevertheless, this is a transient region mostly controlled by the TJ motions and involves additional positive entropy production, which is not accounted in the global Helmholtz free energy as presented above. A careful inspection of Fig. 5(c) may show that the surface free energy slowly deviates from linearity by making a turn towards the stationary non-equilibrium state region, and eventually it may be stabilized. This event is closely correlated with the behavior of the base line extension in Fig. 5(d). It is clear that this experiment prematurely terminated before the system reaches to the stationary non-equilibrium state, which is indicated by the plateau regions in the kinetic parameters such as the base extension, the TJ contact angle, and finally the peak height. The reason for this rather premature termination was the need for excessive computation time and memory, otherwise we may get a profile having little more flattened tails. In this experiment, the peak height and the peak halve-width are found to be, respectively, $h_p \cong 1.47 \rightarrow 7.12 \; nm$, and $\bar{w} \cong 2.20 \rightarrow 10.65 \; nm$ for Ge/Si(100) system.

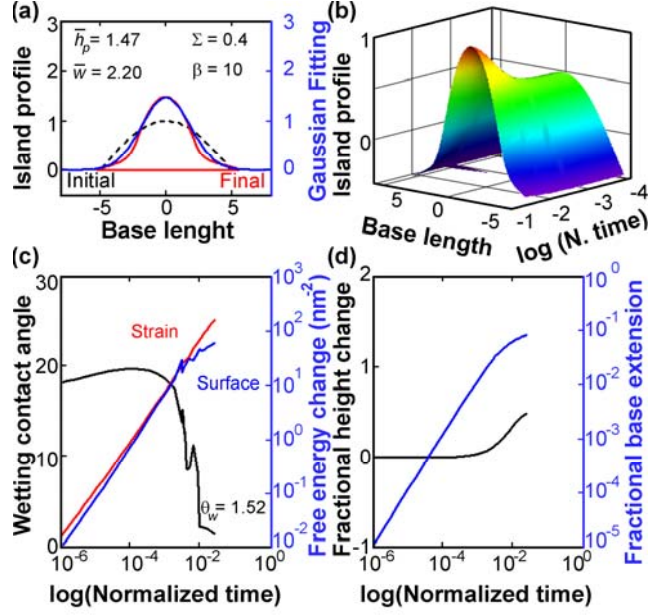

**FIG. 5. (Color online)** (a) The stationary island profile at the transient stage just before the onset of the SK island formation regime. Gaussian bell-shape curve representing the final profile has the following parameters: $\bar{h}_p = 1.47$, $\bar{w} = 2.20$ and thus the peak height-to-peak width ratio of $\xi = 0.334$. (b) 3D time evolution of island profile. (c) Evolution of the contact angle is shown on the left *y*-axis. On the right *y*-axis, the strain energy and surface free energy changes are given for Ge/Si(100) system. (d) Time evolution of peak height and TJ displacement. Simulation Data: $\Sigma = 0.4$, $\bar{h}_p = 1$, $\beta = 10$, $\nu = 0.273$, $\bar{M}_{TJ} = 2$, $\lambda = 1$, $\bar{\delta} = 0.005$, $f_s = 1.2$ and $f_d = 1$.

In Fig. 6, we illustrate the effect of decrease in the aspect ratio on the threshold level of ESED for the formation of SK islands, which shows a substantial increase in ESED parameter from $\Sigma = 0.30$ for $\beta = 28$ to $\Sigma = 0.50$ for $\beta = 10$. Our findings on the stationary values, which describes the morphology of SK in terms of a fourth degree Gaussian profile, may be summarized as: $\bar{h}_{\max} = 1.76$, $\bar{w} = 1.80$ and $\bar{L} = 10.784$. These parameters may be converted into the real space by employing the length scale, $\ell_o \equiv \Sigma \ell^* \to 6.06\ nm$, obtained for the Ge/Si(100) system. This conversion results a peak height of $h_{max} = 10.49\ nm$, halve-peak width of $w = 10.82\ nm$ and the extended domain length of $L = \lambda \simeq 65.30\ nm$ for the SK island formed during the evolution of the droplet having integrated thickness of $h_o \simeq 3.85\ nm$, and the base length of $L = 60.55\ nm$ (i.e., *the original area* $A \cong 233.18\ nm^2$). At the stationary non-equilibrium state the stationary height-to-base length aspect ratio becomes $\beta_s = 5.93$ instead of $\beta = 10$.

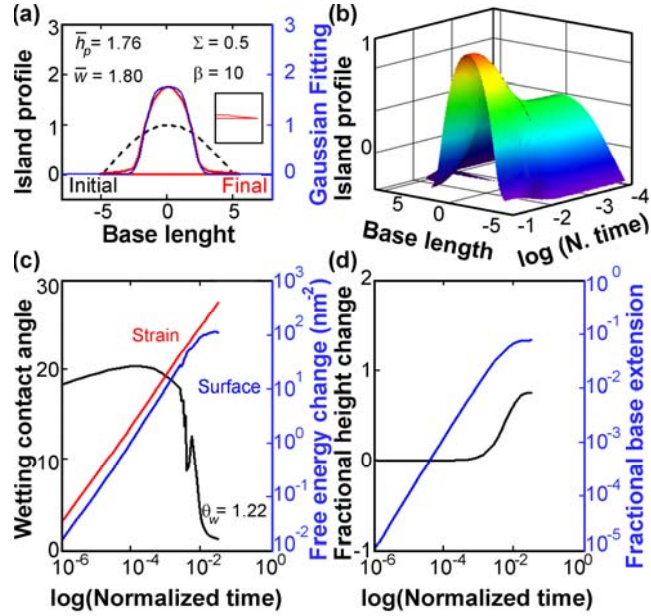

**FIG. 6. (Color online)** (a) Spontaneous formation of the SK island with an almost perfect flat wetting layer from a single crystal. The inset details the structure of the wetting layer at the TJ. The singlet has fourth degree Gaussian bell-shape curve having the following parameters: $\bar{h}_p = 1.76$, $\bar{w} = 1.80$ and thus the peak height-to-peak width ratio of $\xi = 0.489$. (b) 3D time evolution of island profile. (c) Evolution of the contact angle is shown on the left *y*-axis. On the right *y*-axis, the strain energy and surface free energy changes are given for Ge/Si(100) system. (d) Time evolution of peak height and TJ displacement. Simulation Data: $\Sigma = 0.50$, $\bar{h}_p = 1$, $\beta = 10$, $\nu = 0.273$, $\bar{M}_{TJ} = 2$, $\lambda = 1$, $\bar{\delta} = 0.005$, $f_s = 1.2$ and $f_d = 1$.

A careful examination of Fig. 6(c and d) clearly shows that this experiment is also prematurely interrupted at the onset of the stationary non-equilibrium state due to the same computational requirements. Even though the kinetic parameters such as the base line extension, and the wetting angle indicate that they have reached the stationary non-equilibrium state region, the global Helmholtz free energy change still does not show any sign reversal. This situation is closely correlated with Fig. 6(a), where one does not see any well developed flat wetting layer formation compared to its counterpart in Fig. 3(a). The case reported in Fig. 3(c) also shows different kinetic behavior even though topologically both SK islands appear to be very similar, with the exception of the depth and extend of the wetting layers.

In order to correlate two different SK states having exactly the same size in real space, we also performed a special test run using an ESED parameter of $\Sigma = 0.30$, which corresponds to the onset of the SK island formation regime, where the droplet has an aspect ratio of $\beta = 28$, and the normalized peak height of $\bar{h}_p = 1$. These figures in real space match up to a droplet having an initial integrated thickness of $h_o \simeq 2.31\ nm$, and base length of $L = 101.72\ nm$ (i.e., *the original area* $A \cong 235.18\ nm^2$). This test run resulted following output data for the stationary state, which exhibits a fourth degree Gaussian profile: $\bar{h}_{\max} = 2.85$ (peak height) and $\bar{w} = 3.0$ (halve peak width), and $\bar{L} \simeq 10.784$ (extended domain size). In the real space, for the Ge/Si(100) system, these data amounts to: $h_{\max} = 10.35\ nm$, $w = 10.96\ nm$, and $L \equiv \lambda = 101.72\ nm$ with a stationary aspect ratio of $\beta_S = 9.82$.

This is a very interesting result, and clearly shows that two droplets having two different initial shapes, characterized by the two different aspect ratios in the normalized space, but having exactly the same sizes (i.e., area in 2D space) in real space evolved into the SK islands having almost exactly the same shape and size. The only difference between these two systems is in the extensions of the wetting layer platforms, which are defined by the original domain sizes with slight enlargements due the TJ activities. This behavior may be summarized by an analytical expression for the adopted Cosine-shape droplet by writing:

$$A = \frac{2}{\pi}\beta \ell_o^2 \implies \frac{A}{\ell^{*2}} = \frac{2}{\pi}\beta\, \Sigma_\beta^2 \cong 1.59 \tag{12}$$

In Eq. (12), the subscript $\beta$ attached to the ESED parameter, $\Sigma$, and the numeric value of 1.59 indicates the onset value for the appearance of the SK island formation regime (i.e., $\Sigma_{10} \to 0.50$; $\Sigma_{28} \to 0.30$), which may have well defined range or band structure for the singlet and doublet, etc. depending upon the height-to-base length aspect ratio of the droplet. Kukta and Freund[22] found a parabolic connection between the aspect ratio, which defines the shape of the equilibrium island, and the normalized island area: $A/\ell^{*2}$. Their aspect ratio is completely different than ours, and it relies on the ratio of the height-to-base width of the island, which is obtained by a numerical searching technique that is also based on the Cosine-shape initial film morphology, but it is nothing to do with self-evolution of the system towards the stationary non-equilibrium states.

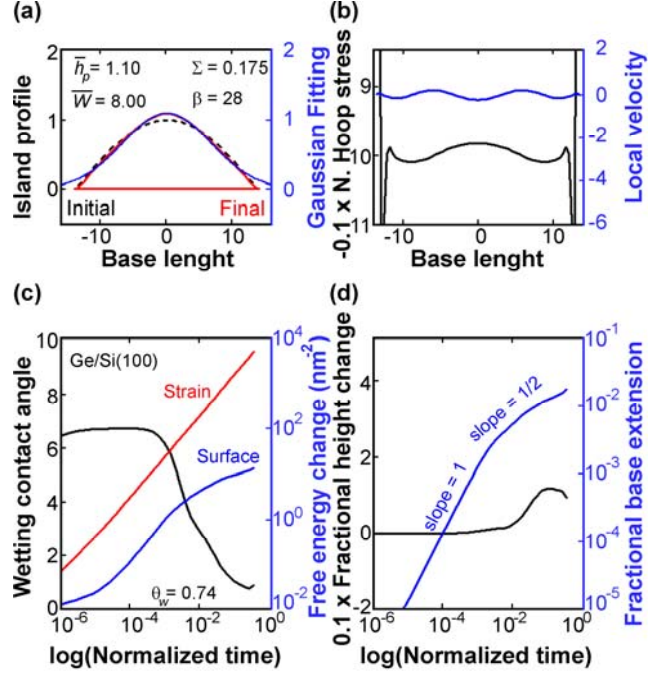

**FIG. 7. (Color online)** (a) Evolution of the Cosine-shape droplets towards the stationary equilibrium state by readjustment of the base length trough the TJ motion. Gaussian bell-shape curve representing the final profile has the following parameters: $\bar{h}_p = 1.10$, $\bar{w} = 8.00$ and thus the peak height-to-peak width ratio of $\xi = 0.069$. (b) Instantaneous velocity and the hoop stress distributions along the final droplet profile. (c) Evolution of the contact angle is shown on the left *y*-axis. On the right *y*-axis, the strain energy and surface free energy changes are given for Ge/Si(100) system. (d) Time evolution of peak height and TJ displacement. Simulation Data: $\Sigma = 0.175$, $\bar{h}_p = 1$, $\beta = 10$, $v = 0.273$, $\bar{M}_{TJ} = 2$, $\lambda = 1$, $\bar{\delta} = 0.005$, $f_s = 1.2$ and $f_d = 1$.

In Fig. 7, the results of a computer experiment, which is executed by using a relatively low value for the ESED parameter (i.e, $\Sigma = 0.175$), are presented. In the case of Ge/Si(100) system, this value for the ESED represents a droplet having a peak height of $h_p = 2.12\ nm$ and the base length of $L = 59.34\ nm$, which may be described by a height-to-width aspect ratio of $\xi = 0.036$, and the normalized area of $A/\ell^{*2} = 0.546$. The profile of this island looks very similar to those described by Kukta and Freund[22] (see Fig. 2 in Ref. [22]) in their remarkable work on the equilibrium island shapes for very small size droplets. The wetting layers at the domain edges are very narrow and about 1.04 *nm*. This tiny droplet as may be deduced from Fig. 7(c and d) is stabilized spontaneously by small adjustments in the base length as well as in the wetting contact angle by the TJ motion. At the start, TJ displacement is linear with time and then turns to a new regime where it demonstrates a new slope of ½ as may be seen from the double logarithmic plot in Fig. 7(d). The TJ has a constant velocity up to knee point then slows down by showing a connection such as $V_{TJ} \sim 1/\sqrt{t}$ up to the onset of the stationary state regime, then levels off. The calculated value of the integrated thickness is $h_o \cong 1.35\ nm$, which is greater than the critical film thickness calculated previously as $h_c^{Ge} = 0.56\ nm$. These values results $h_o / h_c^{Ge} \cong 2.428$, and the growth rate versus film thickness plot for this ratio is given in Fig. 1 for demonstration. According to the linear theory presented previously, the droplet should be in the instability regime, on the other hand this experiment shows that the system is evolving towards the stationary non-equilibrium state with a large and negative global Helmholtz free energy release.

**b. Morphological evolution of droplet with growth**

To show the effect of growth on the morphological evolution of droplet, we performed an experiment using the same input data as it was employed previously to obtain Fig. 3. But this time, we fully considered the growth term in the governing equation (1) by employing the following values for the growth mobility and the Helmholtz free energy of condensation, respectively, $\bar{M}_b = 1$ and $\Delta \bar{F}_{vd}^o = 2$.

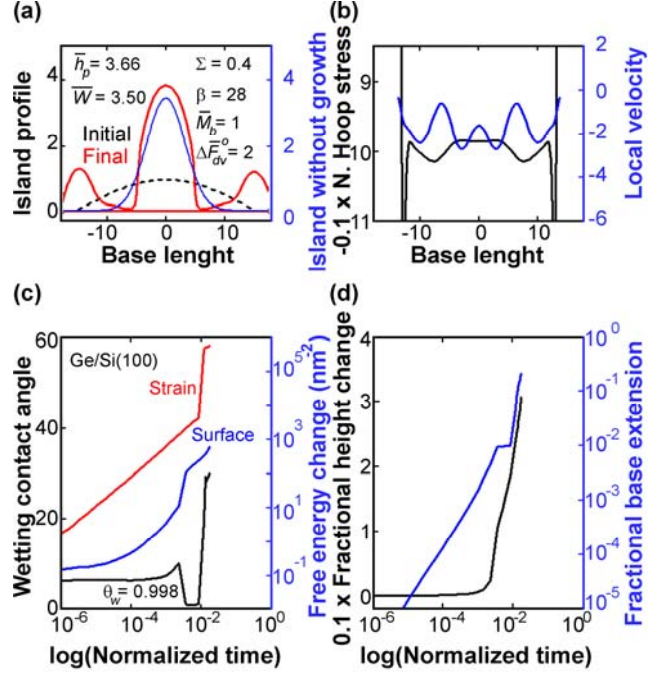

**FIG. 8 (Color online)** (a) The effect of the growth on the SK island morphology: formation of two satellites shouldering the primary pea, and separated by almost perfectly flat wetting layers. Gaussian bell-shape curve representing the final profile has the following parameters: $\bar{h}_p = 1.10$, $\bar{w} = 8.00$ and thus the peak height-to-peak width ratio of $\xi = 0.069$. (b) Instantaneous velocity and the hoop stress distributions along the final droplet profile. (c) Evolution of the contact angle is shown on the left *y*-axis. On the right *y*-axis, the strain energy and surface free energy changes are given for Ge/Si(100) system. (d) Time evolution of peak height and TJ displacement. Simulation Data: $\Sigma = 0.40$, $\bar{h}_p = 1$, $\beta = 28$, $\nu = 0.273$, $\bar{M}_{TJ} = 2$, $\lambda = 1$, $\bar{\delta} = 0.005$, $f_s = 1.2$ and $f_d = 1$ and the growth parameters $\bar{M}_b = 1$, $\Delta \bar{F}_{vd}^o = 2$.

Fig. 8(a) illustrates two profiles with (red) and without (blue) the growth term. In fact, the no growth case was already discussed in the case presented in Fig. 3. In the growth case, we observed a primary peak at the center accompanied by two subsidiary or satellite peaks in each side which altogether covers the computation domain. By zooming this figure, one observes very narrow and thin wetting layers ($\delta \bar{h}_{Ge} \cong 0.0587 \to 0.28\ nm$) separating the satellites from the primary peak. This clearly indicates that we are still in the domain of the SK islands formation regime. As can be seen from the kinetics data presented in Fig. 8(c and d), this system shows some intermediate stationary non-equilibrium state for the time interval of $\bar{t} \approx \{0.05-0.1\}$, where the wetting contact angle $\theta_w \approx 0.998^o$ as well as the size of the computation domain $\delta L/L_o \approx 0.01$ seem to be stabilized as clearly indicated by the appearance of the plateau regions. Similarly, up to the onset of this rather short living intermediate regime, the height of the primary peak does not show any appreciable increase. Otherwise, the system there on evolving continually unless one turn-off the condensation process completely. In Fig. 8(b), the instantaneous velocity and the hoop stress distributions are plotted with respect to the position of the collocation points along the droplet surface. The normalized hoop stress is compressive in sign, since we have had the Ge/Si(100) system in our mind, which has a negative misfit strain of $\varepsilon_o = -0.042$. One observes very high tension stresses concentrated only at the edges of the interface, where the contact between droplet and the substrate takes place through the TJ, which goes up to the level of $\bar{\sigma} \cong 2.25 \to 13.389\ GPa$, and are not illustrated in this diagram. The velocity diagram (Fig. 8(b)) shows two positive maxima, which correspond to the shoulders of the satellites next to the primary peak sides. This indicates that there is a high rate of shrinkage or flatting taking place there, which causes not only the better development of the satellites by rounding off but also the enlargement of the wetting layers next to the primary peak. The velocity distribution shows plateau regions with zero growth rate at the wetting layers, which indicates the stabilization there. Unfortunately, system turned-off automatically because of lag of enough memory space, and we could not pursue further to see the final destiny of the satellites regions. However, one may speculate that those two satellite peaks after a long run time might turn into bell shape subsidiary SK islands extending towards the substrate having very narrow wetting layer with almost zero contact angle.

## IV. CONCLUSIONS

In this paper, we applied the physico-mathematical model, developed by Ogurtani based on the irreversible thermodynamics treatment of surfaces and interfaces with singularities,[23] to describe the dynamical and spontaneous evolution of flat solid droplets (bumps) driven by the surface drift diffusion induced by capillary forces and mismatch stresses, during the development of the Stranski-Krastanow island morphology on a rigid substrate. The present study showed great potential to shade some more lights on the fundamental roles played by those parameters, which describe the shape $\beta$ and the size $\Sigma$ of epitaxially grown droplets, in SK island formation. These parameters, as demonstrated here, dictate selectively what type of SK island would be formed among a large pool of different possibilities (i.e., singlet, doublet, etc.), by the spontaneous evolutions of this isochoric system without having exposed to any external and/or internal perturbations. We also demonstrated that for a given *aspect number, $\beta$*, defined as the height-to-length ratio of the droplet, any desired number of SK island multiples formation may be realized if the strain energy density parameter $\Sigma$ belongs to the well defined closed (bonded) and continuous set of real numbers in the normalized and scaled length-time space. We also revealed that the droplets (i.e., furnished by proper sets of shape and size parameters), having exactly the same size, regardless of their initial shapes may evolve spontaneously into the same SK island morphologies (i.e., same size and shape) in real space. The only difference is the extend of the wetting layer platform. The small aspect ratios result in narrow wetting layer platforms than the large aspect ratio constituents. Finally, we disclosed that the linear instability theories, which heavily depend on the small perturbations (i.e., sinusoidal functions or white noise) on the otherwise smooth surfaces of thin films, are not reliable in predicting the spontaneous evolution (i.e., natural changes with positive internal entropy production) of large scale objects.

## ACKNOWLEDGMENTS

The authors thanks William D. Nix of Stanford University and Dick Bedeaux of Norwegian University of Science and Technology, Trondheim for their interest in our theoretical work on surface and interfaces. Thanks are also due Oncu Akyildiz of METU for his valuable comments on the paper. This work was partially supported by the Turkish Scientific and Technological Research Council, TUBITAK through a research Grant No.107M011.